# Aqueous Solutions of Amino Acid Based Ionic Liquids. Dispersion and Structure

**Vitaly V. Chaban** and **Eudes Eterno Fileti**

Instituto de Ciência e Tecnologia, Universidade Federal de São Paulo, 12247-014, São José dos Campos, SP, Brazil

**Abstract**. New ionic liquids (ILs) are continuously introduced involving an increasing number of organic and inorganic ions. Amino acid based ILs (AAILs) represent a specific interest due to their natural origin and, allegedly, low cost. We apply our recently developed force field for imidazolium-based AAILs to investigate structure properties in their aqueous solutions via molecular dynamics (MD) simulations. By reporting cluster analysis, radial distribution functions and spatial distribution functions, we argue that AAIL ions are well dispersed in the aqueous media, irrespective of the AAIL content. Aqueous solutions of AAILs exhibit desirable properties as solvents for chemical engineering. The AAILs in relatively dilute aqueous solutions (10 mol% AAIL) exist as ion pairs, while more concentrated solutions feature certain amount of larger ionic aggregates.

**Introduction**

Molten salts at ambient temperature and pressure, which contain exclusively ionic species, are referenced to as ionic liquids (ILs).[1-20] These liquids are quite different from molecular liquids by properties and molecular-level organization.[11,15,21,22] Furthermore, it is possible to engineer a large number of ILs with a set of just slightly different physical chemical properties (continuous variation of properties). Variation of ionic species is a great advantage of organic salts based on nitrogen containing heterocycles. At the same time, it has presently become clear that perfect IL is a myth. Non-volatility, thermal stability, low viscosity, negligible toxicity to humans, low melting temperature, and high decomposition temperature cannot be combined in a single IL. Therefore, task-specific ILs must be chased for. The ions must be designed specifically to provide a set of desirable properties, when combined together. These properties should remain stable over a temperature range, which is necessary for applications. A variety of ILs has been, thus far, investigated in their binary mixtures with water for electrochemical applications, separation applications, carbon dioxide capture, etc.[2,3,6,8,14-16,23-28]

Green properties of ILs have been discussed in literature for a long time.[20,29,30] However, it has appeared that not all ILs are green and not all ILs are nontoxic to mammals. Many fluorine containing ILs emit toxic gases, such as HF and $POF_3$, upon hydrolysis of certain well-known anions, such as $BF_4^-$ and $PF_6^-$. Due to fatal interactions of molecular fluorine and fluorine containing gases with living creatures, the simplest solution would be to avoid this element in principle. In many cases, such a radical solution is, unfortunately, impossible.

Amino acid-based ILs (AAILs) have been recently introduced.[31] These compounds contain both amino group and a chiral carbon atom. As such, AAILs are ready to act as a designer platform for task-specific ILs. Unlike many other families of ILs, AAILs are reasonably cheap and can be synthesized in large quantities, starting from biological raw-stuff. Additional advantages include biodegradability and biological activity. It is possible to create AAIL out of only amino acids, i.e. both the cation and the anion are based on an amino acid molecule. This is, perhaps, the most exciting possibility on an ever painful way towards green technology. Chemical modifications of carboxyl group (i.e. esterification) and amino group using only environmentally friendly fragments opens definitely bright perspectives.

The field of AAILs still remains rather new. Some research is devoted to carbon dioxide, $CO_2$, capture under various conditions.[32-35] Effect of water on the $CO_2$ capture ability is described in Ref.[34] The $CO_2/N_2$ selectivity of 100 (perfect separation) at the elevated temperature (373 K) was recorded in the case of tetrabutylphosphonium proline-based membrane under dry conditions. Furthermore, the strong water holding ability of tetrabutylphosphonium proline realized large absorption amounts of $CO_2$ and established a large concentration gradient for the $CO_2$-complex across the membrane. Liu and coworkers applied molecular dynamics simulations with empirical potentials to unveil structure of 1-alkyl-3-methylimidazolium glycine. It seems promising that a single glycine anion is able to coordinate more than two imidazole rings, which arrange nearly parallel to one another.[36] Such a strong coordination probably results in an increased shear viscosity, which may preclude certain applications. In the case of long alkyl chains (exceeding four carbon atoms), the latter are aggregated. Atoms in molecule framework was used in a recent density functional theory study to fit the eight amino acid-based anions to the four imidazolium-based cations.[37] In all cases, longer alkyl chain does not foster cation-anion binding, which may be counterintuitive. An interesting correlation between electronic properties and glass transition temperature is discussed.[37] Ionic solvents are increasingly tested in the context of peptide chemistry both for synthetic and analytical purposes.[38] Peptide synthesis is one of the major challenges in today's chemistry, since laborious protection / de-protection steps constitute a severe drawback in the field. AAILs are envisioned as a set of highly perspective solvents for protein processing.

The primary purpose of our present work is to provide accurate insights, with an atomistic resolution, on the effect of water on ionic structure of the three significantly different 1-ethyl-3-methylimidazolium-based AAILs: 1-ethyl-3-methylimidazolium alanine, [emim][ala], 1-ethyl-3-

methylimidazolium methionine, [emim][met], 1-ethyl-3-methylimidazolium tryptophan, [emim][trp]. We consider 10, 20, and 30 mol% solutions of these AAILs in water. These solutions are essentially concentrated. Note the difference in spatial dimensions between the AAIL ion pair and water molecule. We start with a classical radial distribution methodology to explain cation-water, anion-water, and cation-anion coordination at finite temperature, 310 K. This is unlike it could have been done in the case of electronic structure (e.g. density functional theory) investigation, which is, in practice, limited to the zero-temperature case. Entropic factor is expected to play a decisive role in the AAIL-water systems, as well as it is vital in other ionic liquid containing systems.[39] We supplement description of structure by a comprehensive cluster analysis for ion pairs. This allows to understand how well AAILs are dispersed in water at every concentration and in which way this process can be modulated by an amino acid structure. The investigation is fostered by our own recently introduced force field (FF) for these AAILs.[40] This FF accounts for specific cation-anion non-bonded interactions using a straightforward, computationally cheap, but reliable fixed-charge approach. Therefore, reliability of the reported results is justified appropriately. The same approach was successfully used by one of us a few years ago to investigate imidazolium-based IL mixtures with acetonitrile (another wide-spread molecular solvent).[18-20]

**Methodology**

The analysis of structure of the AAIL-water mixtures was performed based on classical molecular dynamics (MD) simulations using non-additive interaction potentials. 1-ethyl-3-methylimidazolium alanine, [emim][ala], 1-ethyl-3-methylimidazolium methionine, [emim][met], 1-ethyl-3-methylimidazolium tryptophan, [emim][trp], were dissolved in water to form 10, 20, and 30 mol% aqueous solutions. These particular AAILs were selected due to possessing significantly different amino acid residues. Indeed, alanine is a small molecule, which is largely hydrophilic.

We expect perfect miscibility of [emim][ala] with water. In turn, methionine is more hydrophobic, while possessing a sulfur (i.e. chalcogen) atom. The sulfur atom is spatially separated from the hydrophilic moiety by the two methylene groups. Tryptophan is a relatively large molecule, which possesses two fused aromatic rings. Although five-membered ring contains a nitrogen atom, it does not alter a hydrophobic nature of the entire radical. All introduced anions were obtained by deprotonating carboxyl groups of the respective amino acid. The major goal of this work is to underline an impact of these amino acid anions on the dispersion and structure of AAILs in water. The list of the simulated systems is provided in Table 1.

**Table 1**: The list of systems simulated in the present work. The quantity of AAIL ion pairs per system was selected with respect to cation and anion sizes. The number of water molecules was selected to obtain the required molar concentration of each solution

| # | AAIL | # ion pairs | # water molecules | # interaction sites | mol% (AAIL) |
|---|------|-------------|-------------------|---------------------|-------------|
| 1 | [emim][ala] | 100 | 900 | 5,800 | 10 |
| 2 | [emim][ala] | 125 | 500 | 5,375 | 20 |
| 3 | [emim][ala] | 150 | 350 | 5,700 | 30 |
| 4 | [emim][met] | 100 | 900 | 6,500 | 10 |
| 5 | [emim][met] | 125 | 500 | 6,250 | 20 |
| 6 | [emim][met] | 150 | 350 | 6,750 | 30 |
| 7 | [emim][trp] | 75 | 675 | 5,400 | 10 |
| 8 | [emim][trp] | 100 | 400 | 5,700 | 20 |
| 9 | [emim][trp] | 120 | 280 | 6,240 | 30 |

The Cartesian coordinates were saved every 1.0 ps for future processing in accordance with relationships of statistical physics. More frequent saving of trajectory components was preliminarily tested, but no systematic accuracy improvement was found. The first 10 ns of the simulation were regarded as equilibration. The subsequent 100 ns of the simulation were used for structure analysis. All systems were simulated in the constant-pressure constant-temperature ensemble, which allows to attain a natural mass density of each model system. The equations of motion were propagated with a time-step of 2.0 fs. Such a relatively large time-step was possible due to constraints imposed on the carbon-hydrogen covalent bonds (instead of a harmonic potential). Note that 2.0 fs is more than ten times smaller than the period of the fastest harmonic bond oscillation in any of these systems.

The electrostatic interactions were simulated using direct Coulomb law up to 1.2 nm of separation between each two interaction sites. The electrostatic interactions beyond 1.2 nm were accounted for by computationally efficient Particle-Mesh-Ewald (PME) method. It is important to use the PME method in the case of ionic systems, since electrostatic energy beyond the cut-off usually contributes 40-60% of the total electrostatic energy. The Lennard-Jones-12-6 interactions were smoothly brought down to zero from 1.1 to 1.2 nm using the classical shifted force technique. The constant temperature, 310 K, was maintained by the Bussi-Donadio-Parrinello velocity rescaling thermostat[41] (with a time constant of 0.1 ps), which provides a correct velocity distribution for a statistical mechanical ensemble. The constant pressure was maintained by Parrinello-Rahman barostat[42] with a time constant of 1.0 ps and a compressibility constant of $4.5 \times 10^{-5}$ bar$^{-1}$. The compressibility constant only determines a genuine rigidity of the barostat. It should not be numerically equal to physical compressibility of the substance to obtain a correct system density.

All molecular dynamics trajectories were propagated using the GROMACS simulation engine.[43] Analysis of structure properties was performed using the supplementary utilities distributed with the GROMACS simulation suite, where applicable. The AAIL ions were placed in cubic periodic MD boxes (Figure 1), whose initial densities were calculated to approximately correspond to ambient pressure at 310 K.

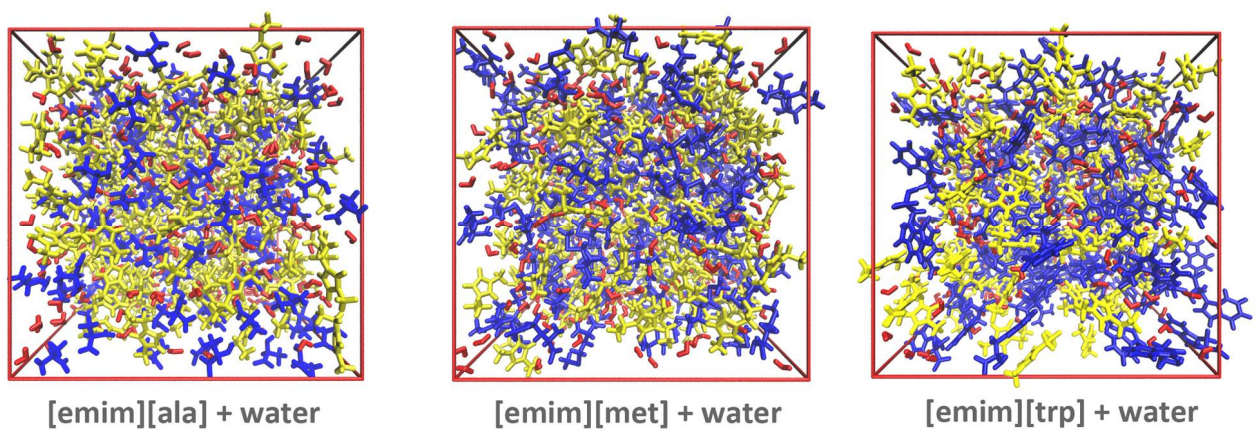

**Figure 1.** Computational unit cells for the three selected amino acid ionic liquids (20 mol%). Imidazolium-based cations are in yellow, amino acid anions are in blue and water molecules are in red.

**Results and Discussion**

Basic description of ion-ionic and ion-molecular structures in liquid matter systems is provided by a set of radial distribution functions (Figures 2-3). All anions are strongly coordinated by surrounded water molecules thanks to carboxyl group (Figure 2). Two sharp and high peaks are positioned at 0.18-0.20 and 0.32-0.36 nm. The position of the first peak univocally suggests formation of a strong anion-solvent hydrogen bond in all systems, irrespective of AAIL content. In turn, it is well-known that imidazolium-based cations exhibit weak to average hydrogen bonding (depending on the external conditions and chemical environment) thanks to their intrinsically acidic hydrogen atom. This hydrogen atom in classical pairwise force fields is denoted by "H5" or "HA" symbols referring to its location near two nitrogen atoms of imidazole ring. The second peak at 0.32-0.36 nm (Figure 2) is smaller (3-4 units) and less sharp. This peak is, however, still high enough to constitute proof of the two well-defined hydration shells of the anion. The third peaks are absent in all RDFs. This feature confirms that even relatively concentrated AAIL solutions in water remain in the liquid state at 310 K. Liquidity of the AAIL solutions is important for applications.

Interestingly, more concentrated solutions possess more ordered structure (both RDF peaks are higher). This is an indirect consequence of slower internal dynamics (transport phenomena) in the AAIL richer systems. Similarly, the RDF peaks in [emim][ala] (10 and 3 units) are smaller than in [emim][met] and [emim][trp] systems (12 and 4 units). Compare the size of [ala] anion with the sizes of [met] and [tpr] anions. As a rule, the anions with larger radicals are less mobile than the anions with smaller radicals.[40] The comprehensive investigation of transport phenomena in the water containing systems is outside the scope of our present work.

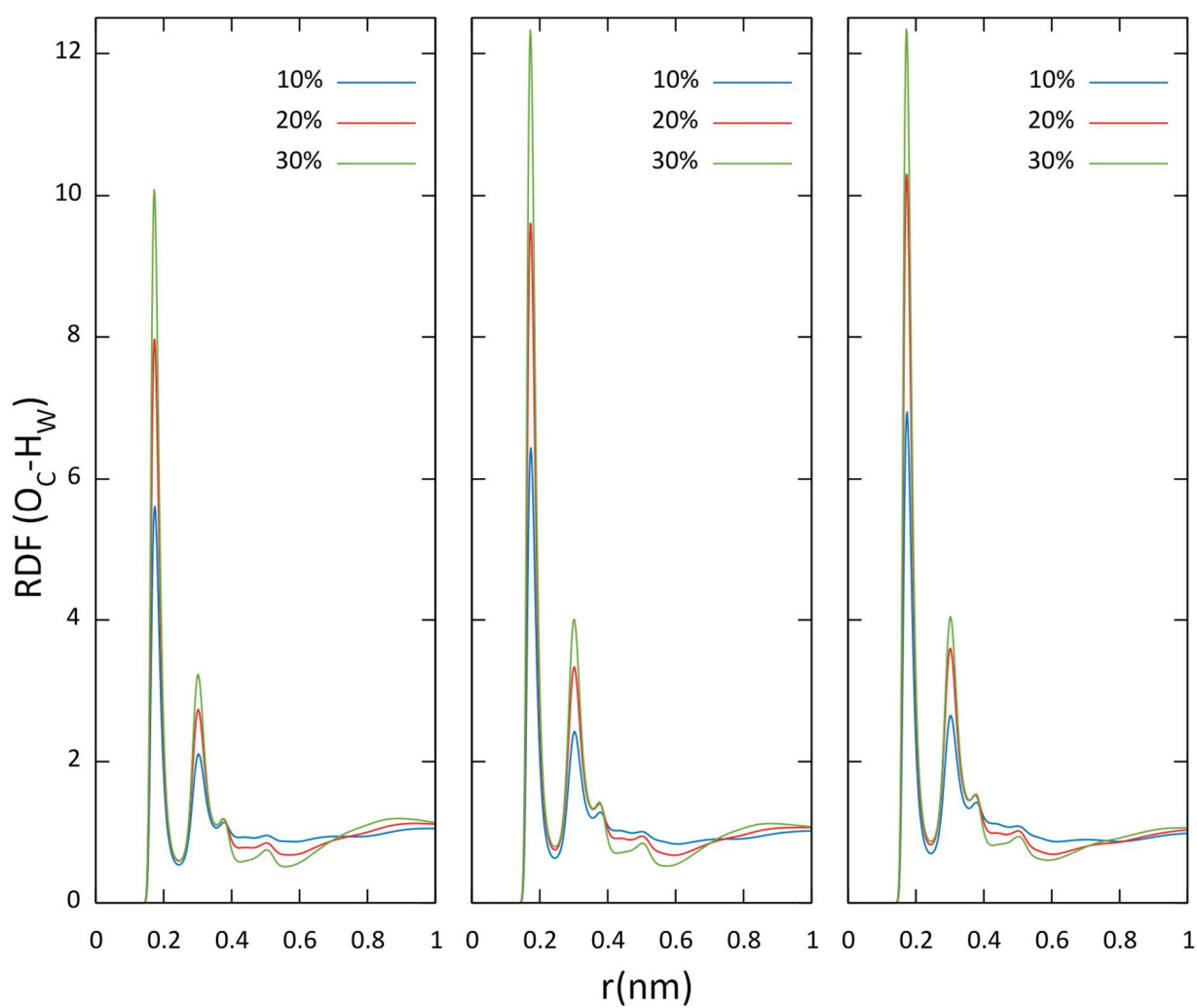

**Figure 2**. Radial distribution functions computed between oxygen atoms of carboxyl group ($O_C$) and hydrogen atoms of water ($H_w$): [emim][ala] (left), [emim][met] (center), [emim][trp] (right). These interaction sites were selected as most electron rich ($O_C$) and electron deficient ($H_W$) atoms in the AAIL-water solutions.

The trend observed in the case of anion-solvent structure persists for cation-anion structure in [emim][ala] and [emim][met], but not in [emim][trp]. Indeed, larger content of AAIL clearly correlates with higher RDF peaks in all systems. The observation, however, appears inverse in the case of [emim][trp]. We hypothesize that due to a larger size of the [trp] anion, its center-of-mass is spatially separated from the cation-anion coordination site (imidazole ring and carboxyl group). Therefore, even though imidazolium ring and carboxyl group exhibit a strong spatial correlation, their centers-of-masses can be relatively disordered. Decrease in the AAIL concentration fosters formation of the ion pairs (as opposed to larger ionic aggregates, Figure 4) and, hence, stronger cation-anion correlations.

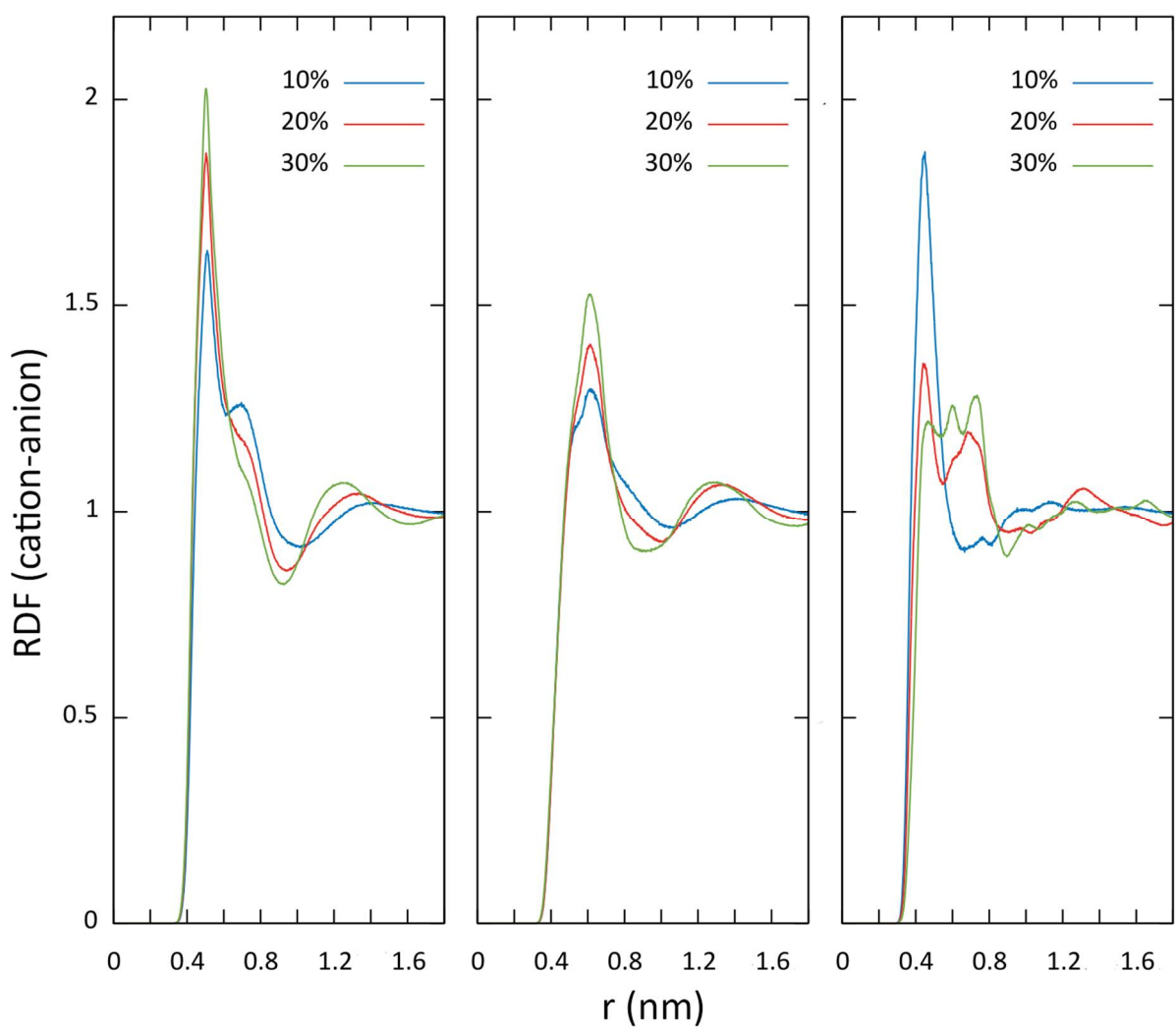

**Figure 3**. Radial distribution functions computed between the center-of-mass of the [emim] cation and the center-of-mass of the amino acid based anion: [emim][ala] (left), [emim][met] (center), [emim][trp] (right). These distributions provide important information regarding average distance between the ions in solution and longer-order ionic structure as a function of ionic concentration.

The major prerequisite of utilization of amino acid based ionic liquids in chemical technology is their miscibility with water and aqueous solutions. The good miscibility would allow to decrease shear viscosity of pure AAILs and to integrate AAILs into biochemically relevant systems. We perform analysis of ionic aggregates in the 10, 20, and 30 mol% aqueous solutions of AAIL (Figure 4). In spite of high concentrations (note the difference in sizes between the solute and the solvent molecules), large ionic aggregates do not exist in any of these systems. The probability of existence of more than six ions within the same aggregate is negligible. This observation indicates a very promising miscibility of water with all three AAILs: [emim][ala], [emim][met], and [emim][trp]. Such a miscibility in all cases was not clear a priori, since large hydrophobic radicals of the AAIL anions might prevent their fine dispersion. Existence of small ionic aggregates in the highly concentrated AAIL solutions favors practical applications of these and less concentrated systems.

The prevailing ionic formation in aqueous AAIL solutions is an ion pair. This pattern takes place in concordance with our recent finding that AAILs feature a single cation-anion coordination site.[40] That is, the ions prefer to coordinate one another using carboxyl group and acidic hydrogen atom of imidazole ring and distribute themselves throughout water volume in the form of neutral cation-anion aggregates. Smaller concentrations favor formation of larger amount of ion pairs, while larger concentrations allow certain percentage of larger aggregates, in addition to ion pairs. Formation of larger aggregates in the concentrated solutions occurs largely due to purely sterical reasons. See Figure 1 for an idea of the volumes, which are occupied by water and AAILs in the 20 mol% solutions.

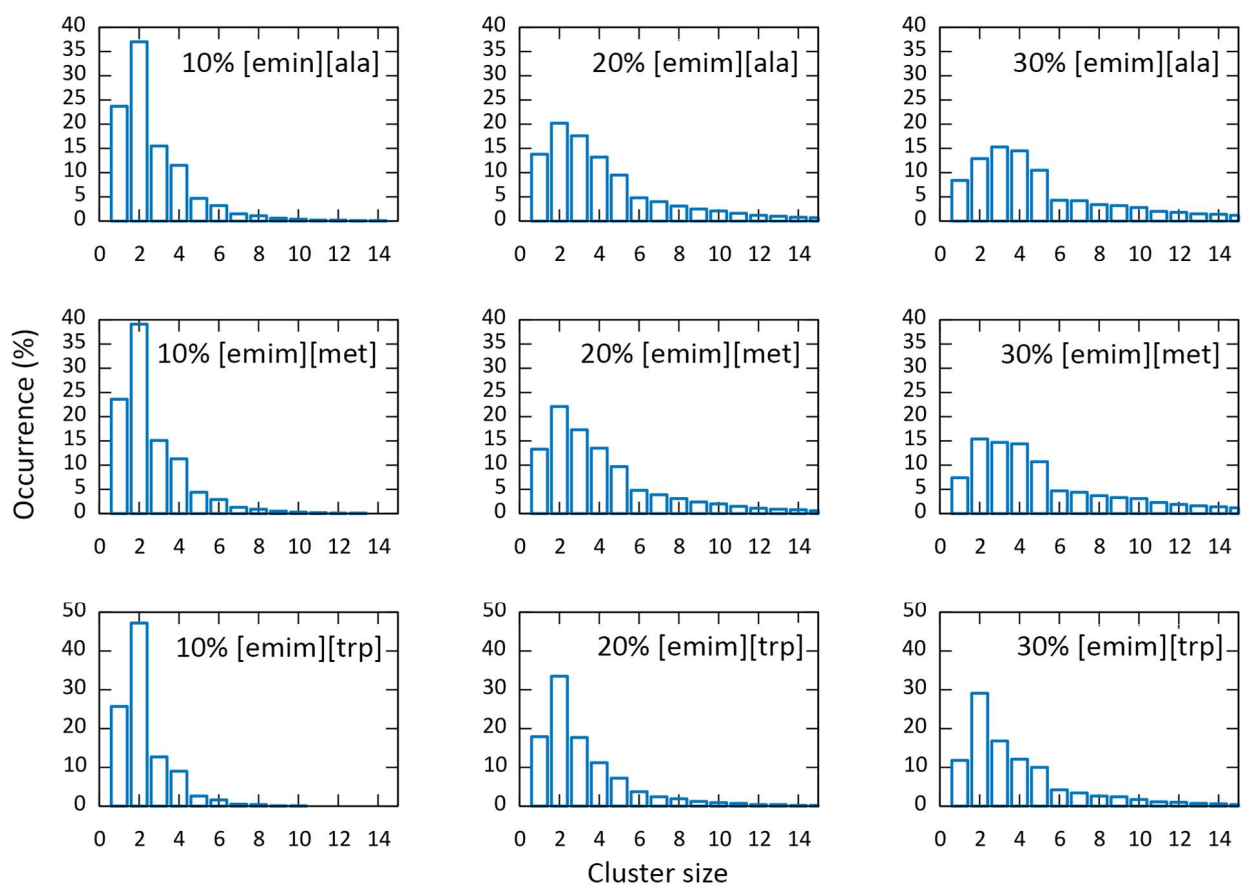

**Figure 4**. Ionic aggregate size distributions in all simulated systems. Ionic aggregates are formed by all AAIL ions, which are present in the systems. See legends for system designations.

Figure 5 unveils the influence of AAIL content on the hydration of the [emim] cation. The [trp] anion is chosen as the most interesting case, as it contains both hydrophilic and hydrophobic moieties. Indeed, spatial distribution of water molecules around the [emim] cation depends on the concentration of AAIL. Whereas the first hydration shell is relatively weak in the 10 mol% solution, it becomes significantly stronger in the 20 and 30 mol% solutions. Furthermore, the 30 mol% solution exhibits the second hydrogen shell, which is sometimes poorly defined in ionic liquids. In all systems, intrinsically acidic "H5" hydrogen atom is well hydrated, while the ethyl chain is not hydrated, irrespective of the AAIL content.

The good hydration of the [emim] cation (Figure 5) coupled with small ionic aggregates (Figure 4) and sharp RDF peaks for anion-water binding (Figure 2) provides a solid evidence that all three AAILs are well dispersed within the volume of water. The solubility limit for AAIL in water, if any exists, is not reached at 30 mol%, which corresponds to very high weight/weight and volume/volume contents of the AAIL.

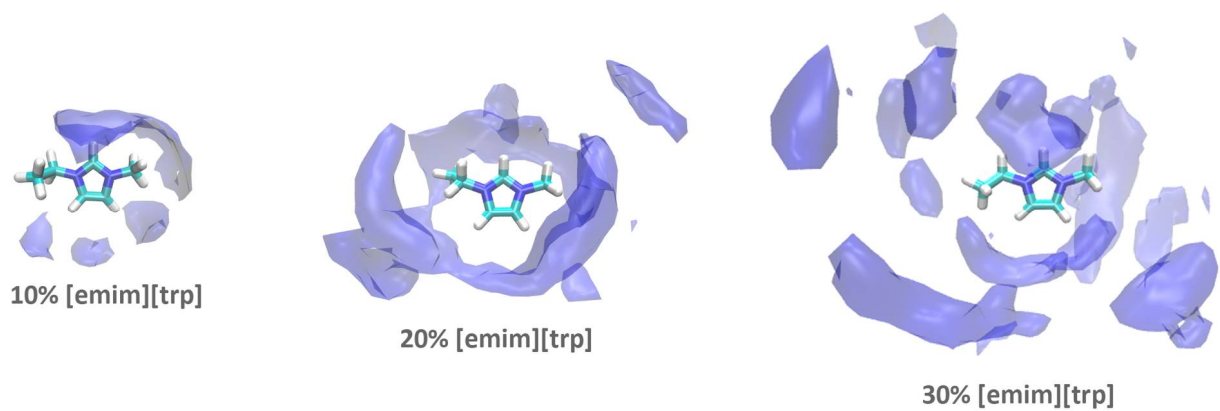

**Figure 5**. Spatial distribution functions of water molecules around the [emim] cation for the three concentrations of [emim][trp] in water. The isovalue of surface is set the same for all three distributions for clear comparison.

**Conclusions**

Classical atomistic-resolution molecular dynamics simulations have been performed for 10, 20, and 30 mol% aqueous solutions of [emim][ala], [emim][met], and [emim][trp] at 310 K. Radial distribution functions, cluster analysis, and spatial distribution of solvent around solute were employed to characterize the structure of the three AAILs in water at the three concentrations. We show that AAIL exhibits a high solubility in water, which does not depend on the size and nature of amino acid radical. Most AAIL ions exist in the form of ion clusters, especially in the 10 mol% solutions. More concentrated solutions, in turn, feature certain amount of larger aggregates, mainly due to sterical reasons. Generally, the percentage of small ionic aggregates may be used as an alternative measure of the solute throughout the solvent media. The results reported here are definitely encouraging for practical applications of the amino acid based ionic liquids.

**Acknowledgments**

V.V.C. acknowledges research grant from CAPES (Coordenação de Aperfeiçoamento de Pessoal de Nível Superior, Brasil) under "Science Without Borders" program. E.E.F. thanks Brazilian agencies FAPESP and CNPq for support.

**AUTHOR INFORMATION**

E-mail addresses for correspondence: vvchaban@gmail.com (V.V.C.); fileti@gmail.com (E.E.F.)

TOC Graphic

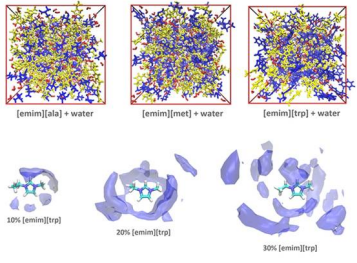